\documentclass[nolinenumbers, RNAAS]{aastex631}
\usepackage{graphicx}
\usepackage{balance}
\usepackage[utf8]{inputenc}
\usepackage{amssymb}
\usepackage{placeins}
\usepackage{natbib}
\usepackage{textcomp}

\usepackage{fontspec}
\usepackage[english]{babel} 
\babelprovide{punjabi} 
\babelfont[punjabi]{rm}[
  Renderer=Harfbuzz,
  Script=Gurmukhi
]{FreeSerif.otf} 

\newcommand{\Punj}[1]{\selectlanguage{punjabi}{#1}\selectlanguage{english}}

\begin{document}

\title{LEGGOS III: Mapping Star Formation and Dust in Gravitationally Lensed Galaxies with \textit{SUMAC}, a UMAP and Clustering Framework}

\author[0009-0005-8103-5823]{Alex Ross}
\affiliation{Department of Astronomy \& the DiRAC Institute, University of Washington, Physics-Astronomy Building, Box 351580, Seattle, WA 98195-1700, USA}

\author[0000-0002-3475-7648]{Gourav Khullar (\Punj{ਗੌਰਵ ਖੁੱਲਰ})}
\affiliation{Department of Astronomy \& the DiRAC Institute, University of Washington, Physics-Astronomy Building, Box 351580, Seattle, WA 98195-1700, USA}
\affiliation{eScience Institute, University of Washington, Physics-Astronomy Building, Box 351580, Seattle, WA 98195-1700, USA}

\author[0000-0001-6251-4988]{Taylor Hutchison}
\affiliation{Astrophysics Science Division, Code 660, NASA Goddard Space Flight Center, 8800 Greenbelt Rd., Greenbelt, MD 20771, USA}
\affiliation{Department of Astronomy, University of Maryland, Baltimore County, MD 21250, USA}
\affiliation{Center for Research and Exploration in Space Science and Technology, NASA/GSFC, Greenbelt, MD 20771 USA}

\author[0009-0000-5333-9970]{Dylan Berry}
\affiliation{Department of Astronomy \& the DiRAC Institute, University of Washington, Physics-Astronomy Building, Box 351580, Seattle, WA 98195-1700, USA}

\author[0000-0002-2525-9647]{Aritra Ghosh}
\affiliation{Department of Astronomy, University of Washington, Physics-Astronomy Building, Box 351580, Seattle, WA 98195-1700, USA}
\affiliation{The DiRAC Institute, University of Washington, Box 351580, Seattle, WA 98195, USA}
\affiliation{eScience Institute, University of Washington, Physics-Astronomy Building, Box 351580, Seattle, WA 98195-1700, USA}

\author[0000-0002-9204-3256]{T.\ Emil Rivera-Thorsen}
\affiliation{The Oskar Klein Centre, Department of Astronomy, Stockholm University, AlbaNova, 10691 Stockholm, Sweden}

\author[0000-0003-3266-2001]{Guillaume Mahler}
\affiliation{STAR Institute, Quartier Agora - All\'ee du six Ao\^ut, 19c B-4000 Li\`ege, Belgium}

\author[0000-0003-1815-0114]{Brian Welch}
\affiliation{International Space Science Institute, Hallerstrasse 6, 3012 Bern, Switzerland}

\author[0009-0004-9243-3459]{Pedram Abedi}
\affiliation{Department of Astronomy, University of Michigan, 1085 S. University Ave, Ann Arbor, MI 48109, USA}

\author[0000-0002-4536-5463]{Cole Panzer}
\affiliation{Department of Physics, University of Cincinnati, Cincinnati, OH 45221, USA}

\author[0000-0001-5097-6755]{Michael Florian}
\affiliation{Eureka Scientific, 2452 Delmer Street Suite 100 Oakland, CA 94602-3017}

\author[0000-0002-5293-3975]{Julissa Sarmiento}
\affiliation{Department of Physics and Astronomy, University of Pittsburgh, Pittsburgh, PA 15260, USA}

\author{the JWST LEGGOS Collaboration}

\correspondingauthor{Alex Ross}
\email{adr55@uw.edu}


\begin{abstract}
Strong gravitational lensing combined with JWST's spatio-spectral resolution enables resolved studies of star-forming regions in $z\sim2$--4 galaxies, but identifying and characterizing such regions in lensed integral-field and multi-band data remains a manual, observer-dependent process. We present \texttt{SUMAC} (Software for the Uniform Manifold Approximation of Clumps), an unsupervised learning pipeline that segments JWST imaging and spectroscopy at the ``spaxel'' level by combining \texttt{UMAP}-based manifold embedding with \texttt{HDBSCAN} density clustering applied to spectral energy distributions/spectra. We demonstrate the pipeline on JWST/NIRSpec PRISM IFS observations of the lensed galaxy SGAS111020.0+645950.8 at $z = 2.481$, recovering six physically distinct stellar/nebular populations. The cluster median SEDs separate cleanly on the presence and strength of H$\beta$+[OIII], H$\alpha$+[NII], $\beta_{NUV}$ slope, Balmer break strength, and the Balmer decrement, with bluer clusters tracing unobscured star-forming regions and progressively redder clusters tracing dusty star-forming regions. 
 
\end{abstract}

\section{Introduction}
The combination of JWST and gravitational lensing can resolve star forming clumps down to $<100\ \mathrm{pc}$
scales even in very distant galaxies \citep[e.g.,][]{rigby2025}. We introduce a novel software pipeline to automate classification of regions in lensed galaxies, \texttt{SUMAC} (Software for the Uniform Manifold Approximation of Clumps), which uses machine learning to analyze photometric or spectroscopic data, reduce dimensionality, and cluster together spaxels with similar spectral energy distribution (SED) shape. \texttt{SUMAC} is optimized for use on the image plane of $z\sim2$--4 lensed galaxies in the LEGGOS survey (Khullar et al. 2026, in prep), but can be extended for use on non-lensed systems. \texttt{SUMAC} is capable of taking NIRSpec (JWST Near Infrared Spectrograph) IFS (integral field spectra) cube data and clustering together spaxels with physically similar continuum and line ratio features (and even photometric clumps from NIRCam). We use ``cluster'' to refer to groups identified by our clustering algorithm in the \texttt{UMAP} embedding, and ``clump'' to refer to spatially contiguous regions of the galaxy with distinct physical properties.

\section{Data and Methods}

We feed \texttt{SUMAC}  NIRSpec IFS PRISM data from the JWST LEGGOS survey (GO-04125 and 3843, PIs: Florian, Khullar, Bayliss) of the $z=2.481$ lensed galaxy SGAS111020.0+645950.8 (SGAS1110). SGAS1110 is a rigorously star-forming galaxy at cosmic noon with a total effective magnification of $\mu_{tot}=24.2_{-1.2}^{+3.4}$ across the primary arc (Abedi et al. 2026 in prep), featuring multiple extreme H$\beta$, H$\alpha$, and [OIII] emitting clumps along the primary arc edge and a high star formation surface density indicative of intense clumpy star formation \citep{Johnson_2017}. This makes it an excellent test case for \texttt{SUMAC's} ability to cluster on discrete physical conditions while reproducing known emission line regions.

\texttt{SUMAC} uses \texttt{UMAP} (Uniform Manifold Approximation and Projection) \citep{mcinnes2020umapuniformmanifoldapproximation} to reduce higher dimensional photometry/spectra to a two-dimensional latent space, similar to \cite{Nielsen_2025} and \cite{Rosito_2023}. \texttt{UMAP} is used in conjunction with the \texttt{HDBSCAN} \citep{McInnes_2017} clustering algorithm to identify regions of data in \texttt{UMAP} space which share similar characteristics. 

First, \texttt{SUMAC} slices an IFS cube into a specified number of bins $n$, to create pseudo-filters ($n$ can be equal to the native IFS cube length as well). The pipeline then applies Gaussian filtering and background subtraction, generating masks based on a threshold of 3$\times$ the background RMS. Each pixel's flux value across all filters is then normalized by dividing by the flux at a common reference wavelength window to isolate spectral shape, which encodes stellar population, ionization, and dust content, from overall brightness, which is set by surface brightness and, for lensed sources, magnification. This ensures that \texttt{SUMAC's} clustering is lensing-agnostic and based on SED-discriminating spectral features. We normalize each spectrum by its median flux in a line-free pseudo-continuum window near rest-frame 4200 \AA. We set \texttt{SUMAC} to train redwards of rest-frame $2900\,\text{\AA}$. 

After normalization, the feature array is passed to the \texttt{UMAP} reducer to collapse the high-dimensional feature array to two dimensions. \texttt{HDBSCAN} then operates on the latent space to identify clusters based on density, assigning each pixel in every distinct cluster a unique color. The now \texttt{HDBSCAN}-colored pixels are mapped back to their original location on the lensed galaxy to identify spatially where unique clusters have been found. For more details on the \texttt{SUMAC} architecture, refer to the \texttt{README} on \href{https://github.com/AlexR055/SUMAC}{Github}.

To test the features that \texttt{SUMAC} identifies and ``clusters'' spaxels into, we compute the median $\beta_{NUV}$ (near-ultraviolet) slope, Balmer break strength (BB), and the Balmer decrement $(\mathrm{H}\alpha + [\mathrm{N\,II}])/(\mathrm{H}\beta + [\mathrm{O\,III}])$ for each \texttt{SUMAC}-defined cluster. We compute the $\beta_{NUV}$ slope by fitting a power law across all median PRISM data points between 2900-3500 \AA\ for each cluster. $\beta_{NUV}$ and Balmer decrement uncertainties are 16th–84th percentiles from bootstrap resampling of member spaxels; BB uncertainties are 16th–84th percentiles of the per-spaxel ratio distribution. Our blue window for the BB is $3450$--$3650\,\text{\AA}$ and the red is $4050$--$4250\,\text{\AA}$. We use the following \texttt{UMAP} configuration for the example shown:
\texttt{n\_neighbors=15}, \texttt{min\_dist=0.1}, \texttt{n\_components=2}, \texttt{metric=euclidean}. For \texttt{HDBSCAN}:
\texttt{min\_cluster\_size}=75, \texttt{min\_samples}=5, \texttt{cluster\_selection\_method}=\texttt{`leaf'}, \texttt{cluster\_selection\_epsilon}=0.48. We select $n=400$ pseudo-filters.

\begin{figure*}[ht!]
    \centering
    \includegraphics[width=\textwidth, height=0.93\textheight, keepaspectratio]{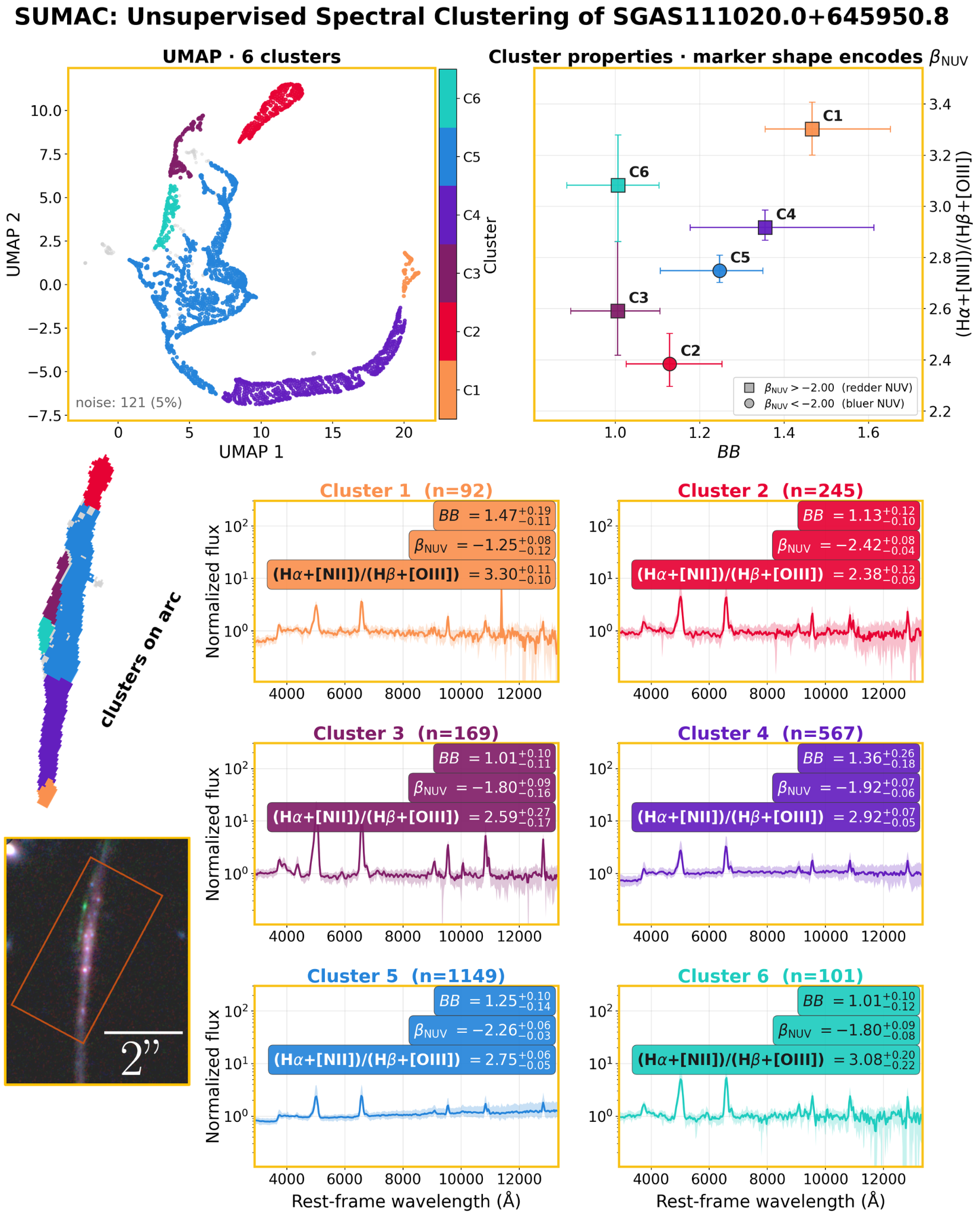}
    \caption{Example of the primary outputs of \texttt{SUMAC}. Shown here is the \texttt{UMAP} embedding of the IFU spaxels (top left) with the \texttt{HDBSCAN} classified and colored spaxel clusters plotted on arbitrary units. The noise points identified by \texttt{HDBSCAN} are gray and do not belong to a specific cluster. In the middle left is the clusters mapped from \texttt{UMAP} space back to their original location on the lensed arc of SGAS1110, and an RGB composite with the region of the galaxy trained on boxed in orange below that. Cluster 3 is coincident with the green emission regions in the image. The six panels below are the median SEDs of each cluster plotted with uncertainty envelopes which span the 16th–84th percentile of the per-spaxel distribution. The top right figure shows the median Balmer decrement plotted against the median Balmer break for each cluster along with the associated bootstrap uncertainty. For an interactive version of this plot, see the \href{https://github.com/AlexR055/SUMAC}{GitHub} repository and this \href{https://youtu.be/u7a9J4rNAOg}{YouTube} video.} 
    \label{fig:main}
\end{figure*}

\section{Results and Discussion}
 \autoref{fig:main} shows that \texttt{SUMAC} identifies six distinct SED clusters within SGAS1110 along with 121 noise points (gray) which do not belong to any cluster. Cluster 1 (C1, orange) has the least negative $\beta_{NUV}=-1.25_{-0.12}^{+0.08}$, the strongest Balmer break $BB=1.47_{-0.11}^{+0.19}$ and the largest Balmer decrement $(\mathrm{H}\alpha + [\mathrm{N\,II}])/(\mathrm{H}\beta + [\mathrm{O\,III}])=3.30_{-0.10}^{+0.11}$. This is indicative of a dust-reddened star forming region with harder ionization conditions, while simultaneously sampling older stars. Cluster 3 (C3, maroon) has the strongest [OIII]+H$\beta$ and H$\alpha$+[NII] emission relative to the other clusters, as well as the weakest Balmer break $BB=1.01_{-0.11}^{+0.10}$ and a more negative $\beta_{NUV}=-1.80_{-0.16}^{+0.09}$. This suggests that C3 is a young, relatively unobscured star-forming clump. Analyzing these two clusters shows that \texttt{SUMAC} has classified two spatially coherent regions of SGAS1110 as star-forming, with the differentiator being one relatively unobscured and one relatively dusty. The top right scatter plot in \autoref{fig:main} shows the median Balmer decrement vs. the median BB for each cluster with associated uncertainties. Each cluster occupies a distinct region in this plot, showing that \texttt{SUMAC} has identified SGAS1110 as possessing numerous distinct physical conditions in its star forming clumps. As such, \texttt{SUMAC} is capable of distinguishing clusters based on subtle variations in line ratios and BB, as well as NUV properties. \texttt{SUMAC} is additionally, with hyperparameter tuning, able to be used as a PSF-sized clump finding tool (in testing, the software segmented clusters 3 and 6 into five smaller clusters, with relative emission strengths and spatial locations matching the findings of Abedi et al. 2026 in prep.)

\texttt{SUMAC} complements the spectro-spatial segmentation paradigm developed for local IFS and unlensed imaging data (\texttt{capivara}, \cite{S_de_Souza_2025}; \texttt{SAGUI}, \cite{desouza2026saguisedbasedsegmentationmultiband}) by extending it to the strongly-lensed cosmic-noon regime, providing a reproducible, automated alternative to manual aperture photometry and clump identification for the LEGGOS sample and other high-redshift lensed-galaxy surveys. \texttt{SUMAC} will be released as an open-source repository on \href{https://github.com/AlexR055/SUMAC}{GitHub}.

\balance
\FloatBarrier

\bibliography{refs}
\end{document}